\documentclass[preprint,5p]{elsarticle}
\journal{Nucl. Instrum. Methods Phys. Res. A}
\usepackage{graphicx}
\usepackage{bm}
\usepackage{hyperref}
\usepackage{booktabs}
\usepackage{amsmath}
\usepackage{amssymb}

\usepackage[separate-uncertainty=true,exponent-product = \cdot,retain-unity-mantissa=false]{siunitx}
\usepackage{enumitem}

\begin{document}

\begin{frontmatter}

\title{Optimizing neutron moderators for a spallation-driven ultracold-neutron source at TRIUMF}

\author[TRIUMF]{W.~Schreyer}
\ead{wschreyer@triumf.ca}
\author[TRIUMF]{C.~A.~Davis}
\author[KEK]{S.~Kawasaki}
\author[Kyoto]{T.~Kikawa}
\author[TRIUMF]{C.~Marshall}
\author[JPARC]{K.~Mishima}
\author[KEK]{T.~Okamura}
\author[TRIUMF,SFU]{R.~Picker}

\address[TRIUMF]{TRIUMF, Vancouver, BC V6T 2A3, Canada}
\address[KEK]{KEK, Tsukuba 305-0801, Japan}
\address[Kyoto]{Kyoto University, Kyoto 606-8501, Japan}
\address[JPARC]{J-PARC, Tokai 319-1195, Japan}
\address[SFU]{Simon Fraser University, Burnaby, BC V5A 1S6, Canada}


\date{\today}

\begin{abstract}
We report on our efforts to optimize the geometry of neutron moderators and converters for the TRIUMF UltraCold Advanced Neutron (TUCAN) source using MCNP simulations. It will use an existing spallation neutron source driven by a \SI{19.3}{\kilo\watt} proton beam delivered by TRIUMF's \SI{520}{\mega\electronvolt} cyclotron. Spallation neutrons will be moderated in heavy water at room temperature and in liquid deuterium at \SI{20}{\kelvin}, and then superthermally converted to ultracold neutrons in superfluid, isotopically purified $^4$He. The helium will be cooled by a $^3$He fridge through a $^3$He-$^4$He heat exchanger.

The optimization took into account a range of engineering and safety requirements and guided the detailed design of the source. The predicted ultracold-neutron density delivered to a typical experiment is maximized for a production volume of \SI{27}{\liter}, achieving a production rate of \SIrange{1.4e7}{1.6e7}{\per\second} with a heat load of \SI{8.1}{\watt}. At that heat load, the fridge can cool the superfluid helium to \SI{1.1}{\kelvin}, resulting in a storage lifetime for ultracold neutrons in the source of about \SI{30}{\second}. The most critical performance parameters are the choice of cold moderator and the volume, thickness, and material of the vessel containing the superfluid helium.

The source is scheduled to be installed in 2021 and will enable the TUCAN collaboration to measure the electric dipole moment of the neutron with a sensitivity of \SI{1e-27}{\elementarycharge\centi\meter}.
\end{abstract}

\begin{keyword}
Ultracold neutrons, moderator, spallation, superfluid helium
\end{keyword}

\end{frontmatter}



\section{Introduction}

If neutrons are slowed below a certain energy, their wavelengths become long enough that they cannot resolve the strong potential of individual nuclei anymore. Instead, materials appear as a constant average potential that can reach several hundred nanoelectronvolts. \cite{Zeldovich} predicted that neutrons with energies below that potential would be totally reflected from such surfaces and could be trapped for long periods of time, making these ``ultracold'' neutrons (UCN) an ideal tool to measure the properties of neutrons with high precision. This phenomenon was first experimentally confirmed by~\cite{Lushchikov,STEYERL196933}.

These early experiments separated ultracold neutrons from the low-energy tail of a thermal neutron spectrum produced in reactors with graphite, water, or polyethylene moderators~\cite{Lushchikov,GROSHEV1971293,ROBSON1972537}. Ultracold neutrons were reflected along a curved tube while neutrons with higher energies can penetrate it and are absorbed in the surrounding material. The first dedicated sources made use of a larger portion of a thermal neutron spectrum by extracting neutrons vertically from a cold moderator~\cite{STEYERL196933,ALTAREV1980413}, reducing their kinetic energy by the gravitational potential of \SI{102.5}{\nano\electronvolt\per\meter}, and eventually reflected them off moving blades mounted on a ``turbine''~\cite{STEYERL1975461,STEYERL1986347} to slow them to ultracold velocities.

The density of ultracold neutrons that these sources can provide is limited by Liouville's theorem. To overcome this limitation, \cite{GOLUB1975133,Golub1983} suggested a ``superthermal'' source. In suitable materials called converters, e.g.~superfluid $^4$He or solid deuterium, neutrons can induce solid-state excitations and lose virtually all their energy and momentum. The inverse process of upscattering can be suppressed if the material is cooled to low enough temperatures. $^4$He does not absorb neutrons at all, making it possible to increase the lifetime of ultracold neutrons to hundreds of seconds if cooled below \SI{1}{\kelvin}.

Despite this obvious advantage of superfluid $^4$He, the first superthermal sources in operation were based on solid deuterium~\cite{SAUNDERS200455,Frei2007,ANGHEL2009272}. Although absorption and upscattering of ultracold neutrons in solid deuterium limits the UCN lifetime to less than \SI{150}{\milli\second}, solid deuterium has a higher UCN-production cross section and can convert a wider range of neutron energies to UCN. Hence, if the solid deuterium is coupled to a large storage volume, the ultracold neutrons are quickly separated from the converter and can be stored for longer times. Such a source can reach a performance similar to a superfluid-helium source, while operating at a more manageable temperature of about \SI{5}{\kelvin}. These sources rely on cold polyethylene~\cite{SAUNDERS200455}, heavy water and solid hydrogen~\cite{Frei2007}, or heavy water and the solid deuterium itself~\cite{ANGHEL2009272} as moderators. Two more reactor sources using solid deuterium are currently in construction, using solid-hydrogen~\cite{Wlokka} and solid-methane moderators~\cite{KOROBKINA2014169}.

To achieve the low temperatures required for a superfluid-helium converter, the first attempts at KEK~\cite{YOSHIKI1994277} converted a cold-neutron beam to ultracold neutrons to reduce the heat load and therefore required no moderators. This concept is now in operation for a source at Institut Laue-Langevin~\cite{PhysRevC.90.015501} but can also be used to produce ultracold neutrons directly in an experiment cell~\cite{Tsentalovich2014,OSHAUGHNESSY2009171}. These sources are limited by the lower intensities of collimated neutron beams, but concepts that could reach production rates similar to the planned source at TRIUMF at significantly lower heat loads have been proposed~\cite{LYCHAGIN201647}.

\cite{PhysRevLett.108.134801} demonstrated that with a $^3$He fridge it is possible to reach the low temperature required for a superfluid-helium source close to a neutron spallation target. This source was later moved to a new spallation target with up to \num{50} times more beam power (\SI{40}{\micro\ampere}, \SI{483}{\mega\electronvolt}, \SI{19.3}{\kilo\watt}) at TRIUMF~\cite{PhysRevC.99.025503,AHMED2019101}. Thanks to a fast kicker magnet~\cite{PhysRevAccelBeams.22.102401}, this spallation target can be irradiated for arbitrary durations, allowing to optimally match the irradiation time to experimental requirements and reducing the heat load whenever no UCN production is required. This feature provides a major advantage over reactor sources, which have to continuously operate at full cooling power, and other spallation sources, which are irradiated with short beam pulses at fixed intervals.

The source at TRIUMF will undergo an upgrade in 2021 to make use of the full beam power available and enable a measurement of the electric dipole moment of the neutron with a sensitivity of \SI{1e-27}{\elementarycharge\centi\meter}. The main limitation to scaling up the beam power is the increasing heat load that has to be removed from the very cold converter. Hence, projects aiming for even higher UCN production in superfluid helium generally plan to operate at higher temperatures~\cite{SEREBROV2011251,doi:10.1063/1.5109879}, at the cost of increased losses that partially negate the gains in UCN production.

This publication describes how we optimized the neutron moderators and converter for the upgraded TRIUMF UltraCold Advanced Neutron (TUCAN) source, taking into account detailed engineering requirements like pressure-vessel thicknesses, safety restrictions, material activation, and biological shielding.

\section{A superfluid $^4$He converter}

\begin{figure}
\centering
\includegraphics[width=\columnwidth]{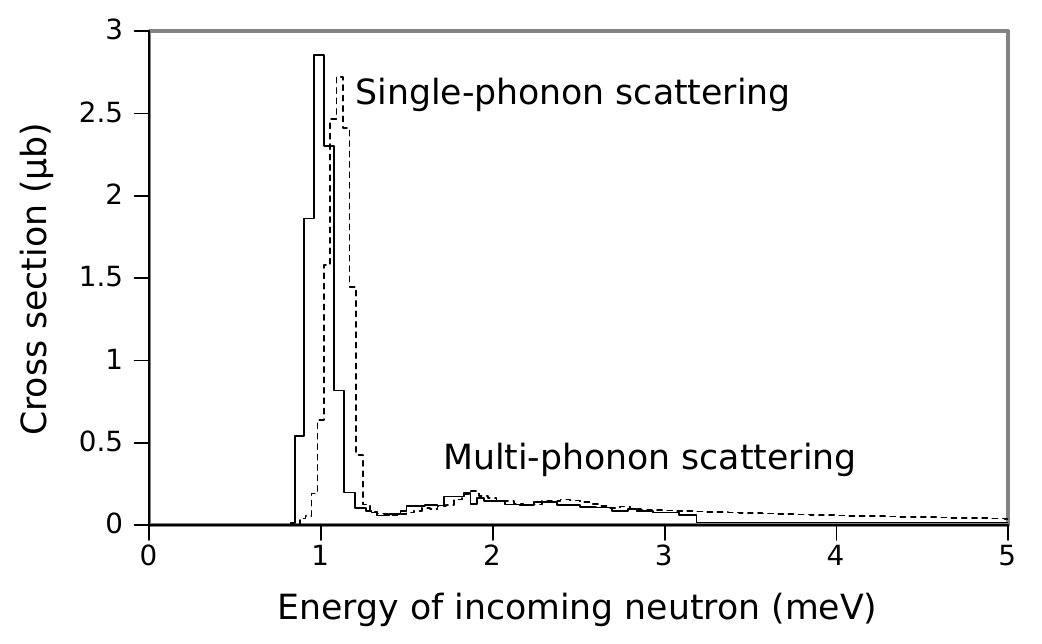}
\caption{Cross sections for production of ultracold neutrons with energies up to \SI{233.5}{\nano\electronvolt} in superfluid $^4$He, calculated from~\cite{PhysRevC.92.024004} (solid line) and~\cite{KOROBKINA2002462} (dashed line).}
\label{fig:UCNprodxs}
\end{figure}

$^4$He in its superfluid state He-II can convert cold neutrons with an energy of \SI{1}{\milli\electronvolt} to ultracold neutrons through single-phonon excitation and, to a smaller degree, through multi-phonon excitation at slightly higher energies~\cite{PhysRevC.92.024004,KOROBKINA2002462}, see Fig.~\ref{fig:UCNprodxs}. To maximize the UCN production, the neutron flux at \SI{1}{\milli\electronvolt} should be maximized by moderating it to an effective neutron temperature of \SI{7.7}{\kelvin}. Cold-neutron sources providing neutrons at these temperatures often use a two-stage moderation first in water and then liquid hydrogen. While hydrogen has a large neutron-scattering cross section, leading to very rapid moderation that is useful for pulsed sources, deuterium has a much smaller neutron-absorption cross section. Hence, heavy water and liquid deuterium are often preferable for continuous sources.

The total loss rate $\tau^{-1}$ of UCN in a volume filled with superfluid helium is given by the sum of all loss channels
\begin{equation}
\tau^{-1} = \tau_\mathrm{He}^{-1} + \tau_\mathrm{abs}^{-1} + \tau_\mathrm{wall}^{-1} + \tau_\beta^{-1}.
\end{equation}
The upscattering loss rate of UCN in superfluid $^4$He, $\tau_\mathrm{He}^{-1}$, is strongly temperature dependent and scales with
\begin{equation}
\tau_\mathrm{He}^{-1} = B \left( \frac{T}{\SI{1}{\kelvin}} \right) ^7,
\label{eq:upscattering}
\end{equation}
with an upscattering parameter $B$ between \SIlist{0.008;0.016}{\per\second}~\cite{GOLUB1975133,PhysRevC.93.025501,PhysRevC.99.025503}. The other stable helium isotope, $^3$He, has an extremely large neutron-absorption cross section. In natural helium with a $^3$He abundance of \numrange{1e-6}{1e-7} the absorption lifetime $\tau_\mathrm{abs}$ is less than \SI{100}{\milli\second}. To reduce the upscattering and absorption losses to a similar level as the losses at the walls $\tau_\mathrm{wall}^{-1} \lesssim (\SI{100}{\second})^{-1}$, the superfluid helium has to be isotopically purified to reduce the $^3$He abundance to below \num{1e-10} and cooled to temperatures of \SI{1}{\kelvin} or less.

With these improvements, the storage lifetime of ultracold neutrons in a source filled with superfluid helium can reach hundreds of seconds. Once UCN production with production rate $P$ is started, ultracold neutrons will accumulate in the source and their density $\rho$ in a volume $V$
\begin{equation}
\rho = \frac{P \tau}{V} \left( 1 - e^{-t/\tau} \right) 
\end{equation}
will reach an equilibrium value $P \tau / V$ after a production time $t$ much longer than the storage lifetime $\tau$.

The ultimate limit is given by the beta-decay lifetime of free neutrons $\tau_\beta = \SI{879.4+-0.6}{\second}$~\cite{PhysRevD.98.030001}.

\section{Simulation model}
\label{sec:simulationmodel}

\begin{figure}
\centering
\includegraphics[width=\columnwidth]{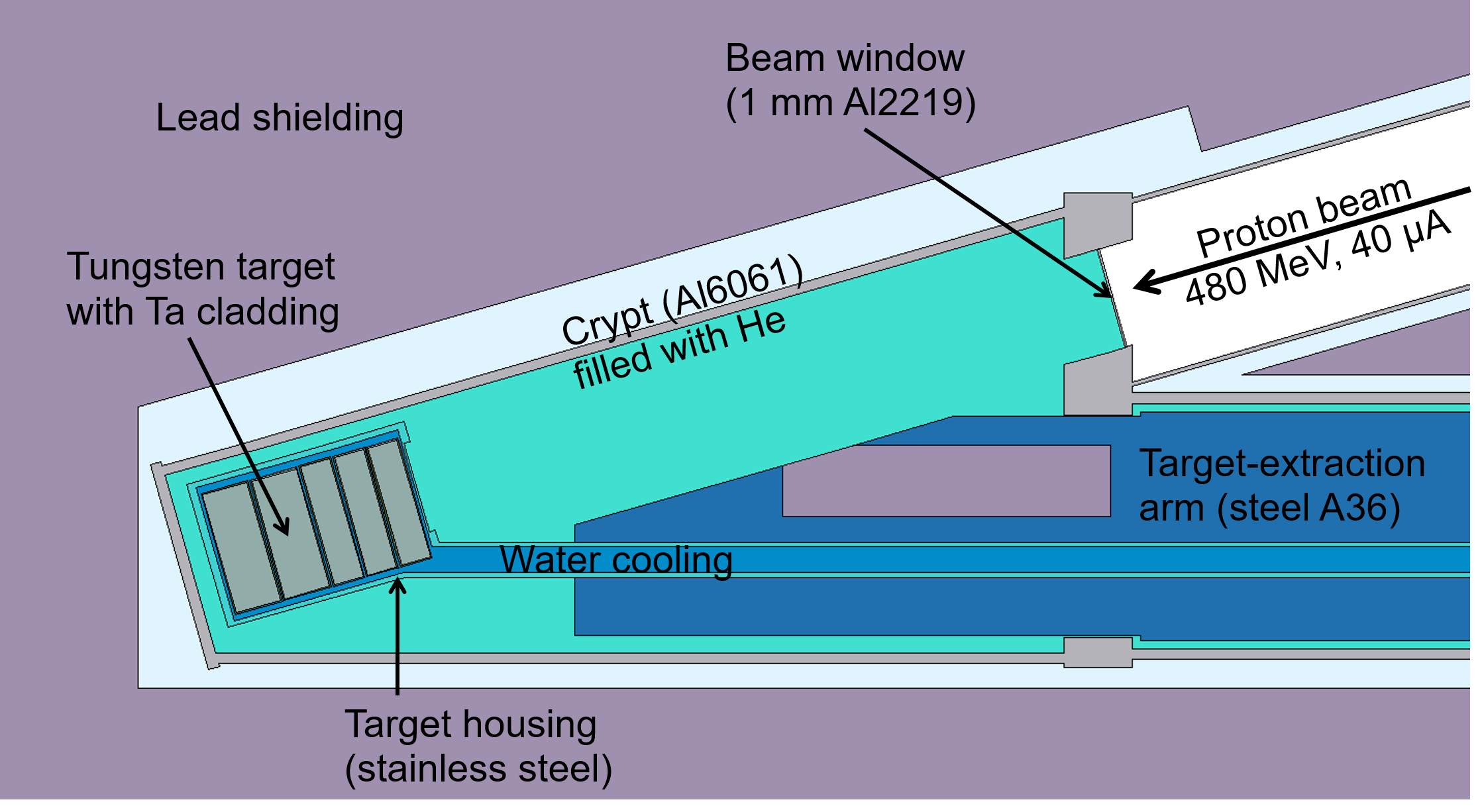}
\caption{MCNP model of the neutron spallation target (top view)}
\label{fig:target}
\end{figure}

To estimate and optimize the performance of the upgraded ultracold-neutron source at TRIUMF, we built a detailed model to simulate it with MCNP 6.1~\cite{doi:10.13182/NT11-135}. The 40~$\mu$A beam of 483~MeV protons from TRIUMF's cyclotron hits the neutron spallation target, see Fig.~\ref{fig:target}, and produces secondary protons, neutrons, electrons, and gamma photons. MCNP simulates the interaction of those particles with surrounding materials with detailed and well-benchmarked scattering models~\cite{CHADWICK20062931}. Many of the materials used in the simulation had realistic compositions determined from assays, see table \ref{tab:materials}. For the other materials, typical compositions were used with the conservative assumption that all impurities are at their maximum allowed levels with natural isotopic abundances.

We benchmarked this simulation model against the prototype source currently in operation at TRIUMF and achieved a good match between the simulated and measured UCN yield~\cite{PhysRevC.99.025503}. However, large uncertainties in the simulations of UCN storage and transport could potentially mask under- or over-estimations in UCN production of as much as \SI{30}{\percent}.

\begin{table*}
\footnotesize
\caption{Material compositions used in the simulation model. Compositions of stainless steel, lead, graphite, and steel shielding were assayed by mass and optical spectrometry~\cite{Sherritt,toyotanso}.}
\centering
\begin{tabular}{l r p{0.7\textwidth}}
\toprule
Material & Density  & Composition (weight-\%)\\
 & (g/cm$^3$) & \\
\midrule
Air & 0.00120 & 75.2 N, 23.2 O, 1.28 Ar, 0.0125 C \\
Al6061 & 2.70 & 95.85 Al, 1.2 Mg, 0.8 Si, 0.7 Fe, 0.4 Cu, 0.35~Cr, 0.25~Zn, 0.15~Mn, 0.15 Ti \\
Al2219 & 2.84 & 91.5 Al, 6.8 Cu, 0.4 Mn, 0.3 Fe, 0.2 Si, 0.15 V, 0.15~Zr, 0.1 Ti, 0.1 Zn, 0.02 Mg \\
AlBeCast 910 \cite{AlBeCast} & 2.17 & 57 Be, 38 Al, 3.4 Ni, 0.5 Si, 0.3 Fe, 0.24 O \\
AlBeMet 162 \cite{AlBeMet} & 2.10 & 62 Be, 38 Al \\
Beryllium & 1.85 & 100 Be \\
Bismuth & 9.75 & 100 Bi \\
Copper & 8.96 & 100 Cu \\
Graphite & 1.70 & 99.978 C, 0.0125 V, 0.0033 Ti, 0.0026 Fe, 0.0014~Al, 0.0006 Ca, 0.0004 Ni, 0.0003~B, 0.0002 K, 0.0002~Si, 0.00016 Zr, 0.0001 Cu, 0.0001~Pb, 0.00006 Zn, 0.00005 Na, 0.00003~Cr, 0.00003~Co, 0.00002 Mg, 0.00002 Mn, 0.00002 Gd, 0.000003~Li \\
Aluminium & 2.70 & 100 Al \\
AZ 80 & 1.80 & 90.85 Mg, 8.5 Al, 0.5 Zn, 0.15 Mn \\
Beralcast 310 \cite{BerAlCast} & 2.16 & 60 Be, 36 Al, 2.5 Ag, 0.25 Si, 0.2~Co, 0.2 Ge, 0.2 Fe \\
Heavy water & 1.10 & 100 $^2$H$_2$O \\
He gas & 0.000180 & 100 He \\
Lead & 11.4 & 99.9915 Pb, 0.004 Bi, 0.001 Cu, 0.0008 Ag, 0.0008~Sb, 0.0005~As, 0.0005 Sn, 0.0005 Fe, 0.0004~Zn \\
LD$_2$ & 0.160 & 100 $^2$H$_2$ \\
Liquid $^3$He & 0.082 & 100 $^3$He \\
Magnox AL80 & 1.80 & 99.2 Mg, 0.8 Al, 0.004 Be \\
Mild steel & 7.80 & 98 Fe, 1 Mn, 0.4 C, 0.28 Si, 0.2 Cu, 0.05 S, 0.04~P \\
Stainless steel & 8.00 & 71.4 Fe, 18 Cr, 8 Ni, 1.32 Mn, 1 Si, 0.095 Co, 0.08~C, 0.045 P, 0.03 S, 0.0075 Nb, 0.0022 As, 0.00036 Sb \\
Steel shielding & 7.36 & 89.15 Fe, 4.5 Cr, 3.3 Ni, 0.99 Cu, 0.8 Mn, 0.35~Si, 0.32~C, 0.25 Al, 0.16 Mo, 0.015 P, 0.057~Co, 0.027~Pb, 0.021 V, 0.02 S, 0.018 Nb, 0.012~Sn, 0.011~Ti, 0.0011 B \\
Superfluid He & 0.145 & 100 $^4$He \\
Tantalum & 16.7 & 100 Ta \\
Tungsten & 19.3 & 100 W \\
\bottomrule
\end{tabular}
\label{tab:materials}
\end{table*}

\subsection{UCN production and heat load}

We determined UCN production $P$ from an MCNP flux tally, multiplying the neutron flux in the UCN converter $\Phi$ with the UCN-production cross section $\sigma$ taken from \cite{PhysRevC.92.024004,KOROBKINA2002462}, see Fig.~\ref{fig:UCNprodxs}:
\begin{equation}
P = \int \Phi(E) \sigma(E) dE.
\end{equation}

\begin{figure}
\centering
\includegraphics[width=0.8\columnwidth]{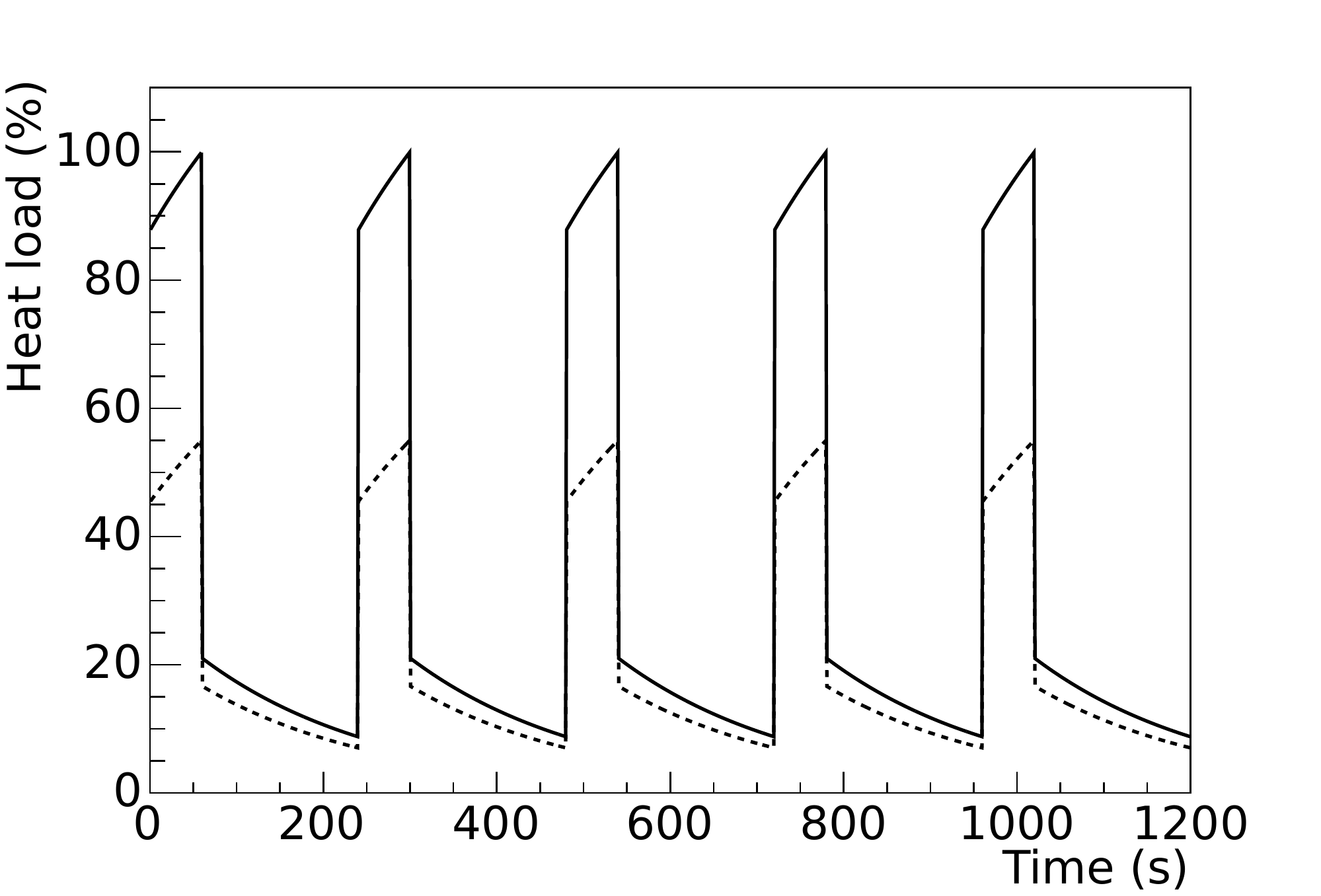}
\caption{Time structure of heat load (solid line) on the converter and its aluminium vessel during irradiation with a duty cycle of \SI{25}{\percent} (beam on for one minute, beam off for three minutes). The dashed line is the fraction deposited in the vessel wall. Static heat load is not included.}
\label{fig:heatload}
\end{figure}

MCNP also provides the average heat deposited per primary proton in each part of the model. Typically, there is a large prompt heat deposited followed by a slowly decaying heat load caused by radioactive decays. To calculate the total heat load during or after a continuous irradiation, this heat-load profile is convolved with the intensity and duty cycle of the irradiation. For these simulations we assumed a duty cycle for a typical UCN experiment that is filled for one minute while the target is irradiated with \SI{40}{\micro\ampere}, and then stops the irradiation while UCN are stored and detected for three minutes, corresponding to a duty cycle of 25~\%, see Fig.~\ref{fig:heatload}.

The heat load calculated in MCNP does not include any static heat load on the cryogenic parts due to thermal conduction, convection, or radiation. For the estimation of the converter temperature we added a static heat load between \SIlist{0;1}{\watt}.

\subsection{UCN-converter temperature and storage lifetime}

\begin{figure}
\centering
\includegraphics[width=\columnwidth]{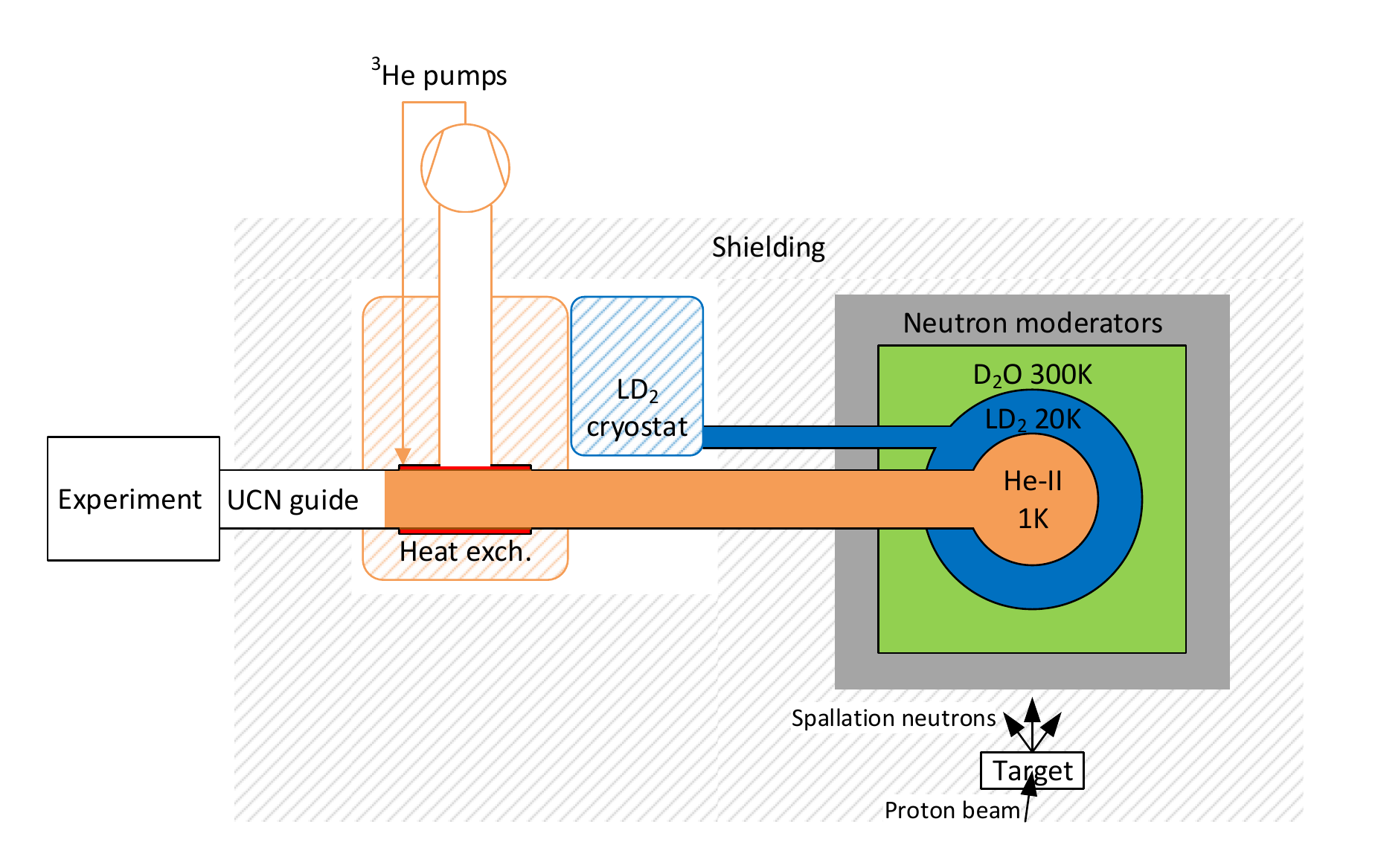}
\caption{Conceptual sketch of the upgraded UCN source at TRIUMF. The proton beam points into the page.}
\label{fig:sketch}
\end{figure}

The converter will be cooled by a $^3$He fridge. A copper heat exchanger filled with about \SI{300}{\milli\liter} of liquid $^3$He is cooled to \SIrange{0.8}{0.9}{\kelvin} by lowering the vapor pressure with an array of vacuum pumps with a pumping speed $S$ of up to \SI{10000}{\cubic\meter\per\hour}. To shield the fridge from the high radiation above the target, a \SI{2.5}{\meter}-long and \SI{150}{\milli\meter}-diameter conduction channel filled with isotopically purified superfluid $^4$He and surrounded by steel shielding connects the heat exchanger to the UCN converter, see Fig.~\ref{fig:sketch}.

To determine the temperature of the superfluid helium we modeled the cooling power of this fridge. The pumps with inlet temperature $T_\mathrm{pump}$ remove heat
\begin{equation}
    Q = \eta_\mathrm{JT}\frac{\left(p(T_\mathrm{^3He}) - \Delta p \right) L S}{R T_\mathrm{pump}}
\end{equation}
from the $^3$He bath with vapor pressure $p(T_\mathrm{^3He})$, latent heat $L$, and universal gas constant $R$. $\eta_\mathrm{JT}$ is the fraction of supplied $^3$He that is converted to liquid during Joule-Thompson expansion and $\Delta p$ the pressure drop in the pumping duct. The same amount of heat has to be transported across the heat exchanger. This heat transport is limited by the Kapitza resistance at the interface areas $A_{^4\mathrm{He}}$ (between liquid $^4$He with temperature $T_{^{4}\mathrm{He}}$ and the copper heat exchanger with temperature $T_\mathrm{Cu}$) given by the Acoustic Mismatch Theory~\cite{van2012helium}:
\begin{equation}
    Q = \SI{20}{\watt\per\square\meter\kelvin\tothe{-4}} \cdot k_G T_\mathrm{Cu}^3 \left| T_\mathrm{Cu} - T_{^{4}\mathrm{He}} \right| A_{^4\mathrm{He}}.
\end{equation}
The Kapitza resistance at the interface area $A_{^3\mathrm{He}}$ between heat exchanger and $^3$He with temperature $T_{^{3}\mathrm{He}}$ is assumed to be \numrange{1.2}{2.6} times higher due to its higher sound velocity:
\begin{equation}
    Q = (7.7 \sim 17) \, \si{\watt\per\square\meter\kelvin\tothe{-4}} \cdot k_G T_\mathrm{Cu}^3 \left| T_\mathrm{Cu} - T_{^{3}\mathrm{He}} \right| A_{^3\mathrm{He}}.
\end{equation}
The scaling factor $k_G$ is determined by the surface quality of the heat exchanger and typically lies between \numlist{20;40}~\cite{van2012helium}. We assume it to be the same for $^3$He. The heat conduction through the copper itself is large enough to be negligible.

\begin{figure}
\centering
\includegraphics[width=\columnwidth]{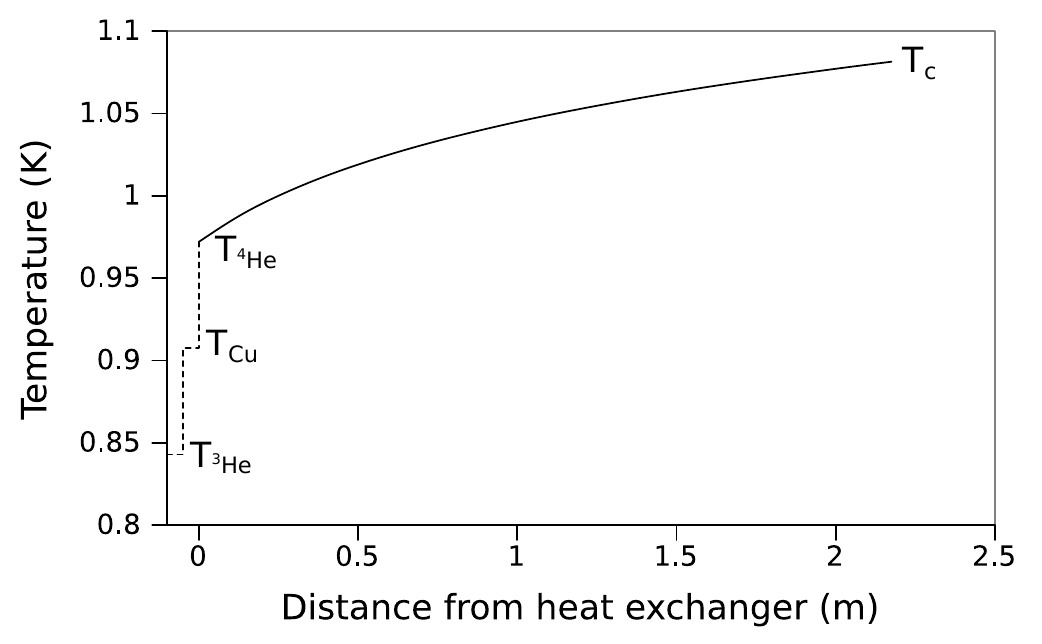}
\caption{Temperature profile in the conduction channel (solid line) and across the heat exchanger (dashed line). This example assumes $\eta_\mathrm{JT} = 0.92$, $\Delta p = \SI{300}{\pascal}$, $S = \SI{10000}{\cubic\meter\per\hour}$, $T_\mathrm{pump} = \SI{300}{\kelvin}$, $k_G = 40$, $A_{^3\mathrm{He}} = \SI{0.76}{\square\meter}$, $A_{^4\mathrm{He}} = \SI{0.24}{\square\meter}$, $Q = \SI{8.1}{\watt} + \SI{1}{\watt}_\mathrm{static}$, and $f(T)^{-1}$ is taken from~\cite{hepak2005}.}
\label{fig:tempprofile}
\end{figure}

The same amount of heat has to again be transported from the converter volume to the heat exchanger through the channel filled with superfluid helium. The heat transported through a channel with length $l$ and cross section $A$ can be determined with the Gorter-Mellink equation~\cite{GorterMellink1949}
\begin{equation}
    Q = \left(\frac{A^3}{l} \int_{T_{^4\mathrm{He}}}^{T_\mathrm{c}} f(T)^{-1} dT \right) ^{1/3},
\end{equation}
with an empirical heat-transport function $f(T)^{-1}$~\cite{van2012helium,hepak2005}. Fig.~\ref{fig:tempprofile} shows an example of the resulting temperature profile along the conduction channel and across the heat exchanger.

\begin{figure}
\centering
\includegraphics[width=\columnwidth]{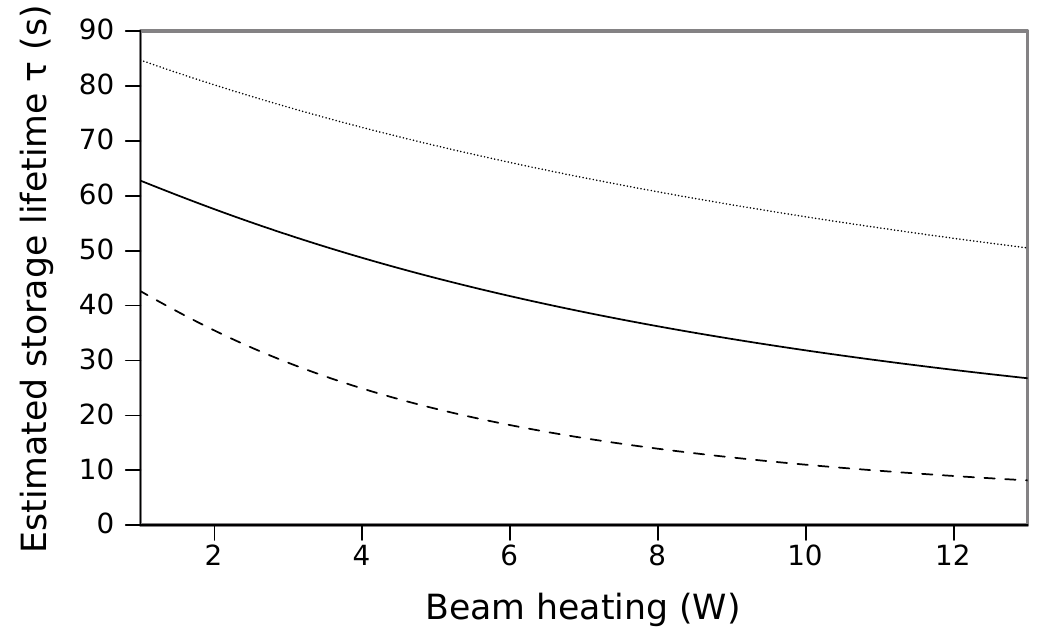}
\caption{Estimated relations between heat load $Q$ and storage lifetime in the source $\tau$ with the most optimistic (dotted line), most pessimistic (dashed line), and an intermediate parameter set (solid line).}
\label{fig:tauVsQ}
\end{figure}

Solving these equations for the converter temperature $T_\mathrm{c}$ gives a relation between $Q$ and $T_\mathrm{c}$ that can be approximated with a power law. Combining it with equation~(\ref{eq:upscattering}) yields an upscattering lifetime of
\begin{equation}
\tau_\mathrm{He} = B \left( \frac{Q}{1~\mathrm{W}} \right)^a
\end{equation}
with $B$ from \SIrange{500}{1500}{\second} and $a$ from \numrange{-1.5}{-1.0}, see Fig.~\ref{fig:tauVsQ}. The heat load $Q$ can also include some static heat load $Q_\mathrm{static}$ due to thermal radiation, convection, and conduction in the cryostat, which we conservatively assumed can be as large as \SI{1}{\watt}.

The converter vessel, conduction channel, and heat exchanger will most likely be coated with NiP. In the NiP-coated prototype source, \cite{PhysRevLett.108.134801} achieved a total storage lifetime $\tau$ of \SI{81}{\second}; our own measurements with a NiP-coated bottle at room temperature with a volume of \SI{6.5}{\liter} and a volume-to-surface ratio of \SI{3.37}{\centi\meter} gave a wall-storage lifetime $\tau_\mathrm{wall}$ of \SI{76}{\second}. Hence, for the optimization, we assumed a range of \SIrange{60}{100}{\second} for $\tau_\mathrm{wall}$. Reaching the required isotopic purity for dozens of liters of helium is difficult with conventional purification methods~\cite{MCCLINTOCK1978201}. Instead we rely on commercially available helium with a $^3$He abundance of \num{1e-12}~\cite{HENDRY1987131,PhysRevC.93.065502}, resulting in an absorption lifetime $\tau_\mathrm{abs}$ of several thousand seconds, so we assumed its contribution to be negligible. See table \ref{tab:assumptions} for the range of assumptions made in the optimizations.

\section{Optimization}

\subsection{Optimization goal}
\label{sec:optimizationgoal}

The primary goal of the TUCAN collaboration is to do a measurement of the neutron electric dipole moment (nEDM) with a sensitivity of \SI{1e-27}{\elementarycharge\centi\meter}. This requires a large number of UCN in the storage cell of the experiment and the source should be designed in such a way that it maximizes the number of UCN that can be delivered to the cell. How many of the produced UCN can actually reach the experiment is largely dominated by losses in the UCN guides between source and experiment, but these losses are not affected by changes in the geometry of the source. If we assume the guide losses are fixed, the source itself has only three parameters that affect the density delivered to the experiment:
\begin{itemize}
\item UCN-production rate $P$,
\item storage lifetime of UCN in the source $\tau$, and
\item volume of the source $V_\mathrm{source}$.
\end{itemize}
To optimize only the source itself, we can use an estimator for the UCN density in the cell $\rho$, given by the total number of UCN produced in the source $P \tau$ diluted into the volume of the whole system $V = V_\mathrm{source} + V_\mathrm{guides} + V_\mathrm{nEDM}$:
\begin{equation}
\rho = \frac{P \tau}{V}.
\label{eq:densityestimator}
\end{equation}
Assuming the UCN guides in the source are \SI{3.0}{\meter} long and have a diameter of \SI{150}{\milli\meter}, and the guides to the experiments are \SI{9.5}{\meter} long and have a diameter of \SI{95}{\milli\meter}, we get $V_\mathrm{guides} = \SI{120}{\liter}$. The two nEDM cells will have a volume of about \SI{30}{\liter} each, totaling $V_\mathrm{guides} + V_\mathrm{EDM} \approx \SI{180}{\liter}$. To account for potential changes of guide diameters and storage experiments with different volumes, we assumed a range between \SIlist{100;200}{\liter}, see table \ref{tab:assumptions}.

\begin{table}
\centering
\caption{Range of assumptions for parameters in the UCN-density estimator $\rho$.}
\begin{tabular}{lr}
\toprule
Parameter & Range \\
\midrule
$\tau_\mathrm{wall}$ & \SIrange{60}{100}{\second} \\
$\tau_\beta$ & \SI{880}{\second} \\
$\tau_\mathrm{abs}$ & $\infty$ \\
$B$ & \SIrange{500}{1500}{\second} \\
$a$ & \numrange{-1.5}{-1.0} \\
$V_\mathrm{guides} + V_\mathrm{EDM}$ & \SIrange{100}{200}{\liter} \\
$Q_\mathrm{static}$ & \SIrange{0}{1}{\watt} \\
\bottomrule
\end{tabular}
\label{tab:assumptions}
\end{table}

During the initial parameter studies we kept the dimensions of the UCN-converter volume constant. Since in that case the factor $V$ in the density estimator $\frac{P \tau}{V}$ is constant and the storage lifetime $\tau$ depends only on heat load $Q$, we can instead optimize the production-to-heat ratio $P/Q$.

\subsection{Initial parameter studies}

\begin{figure}
\centering
\includegraphics[height=3.5cm]{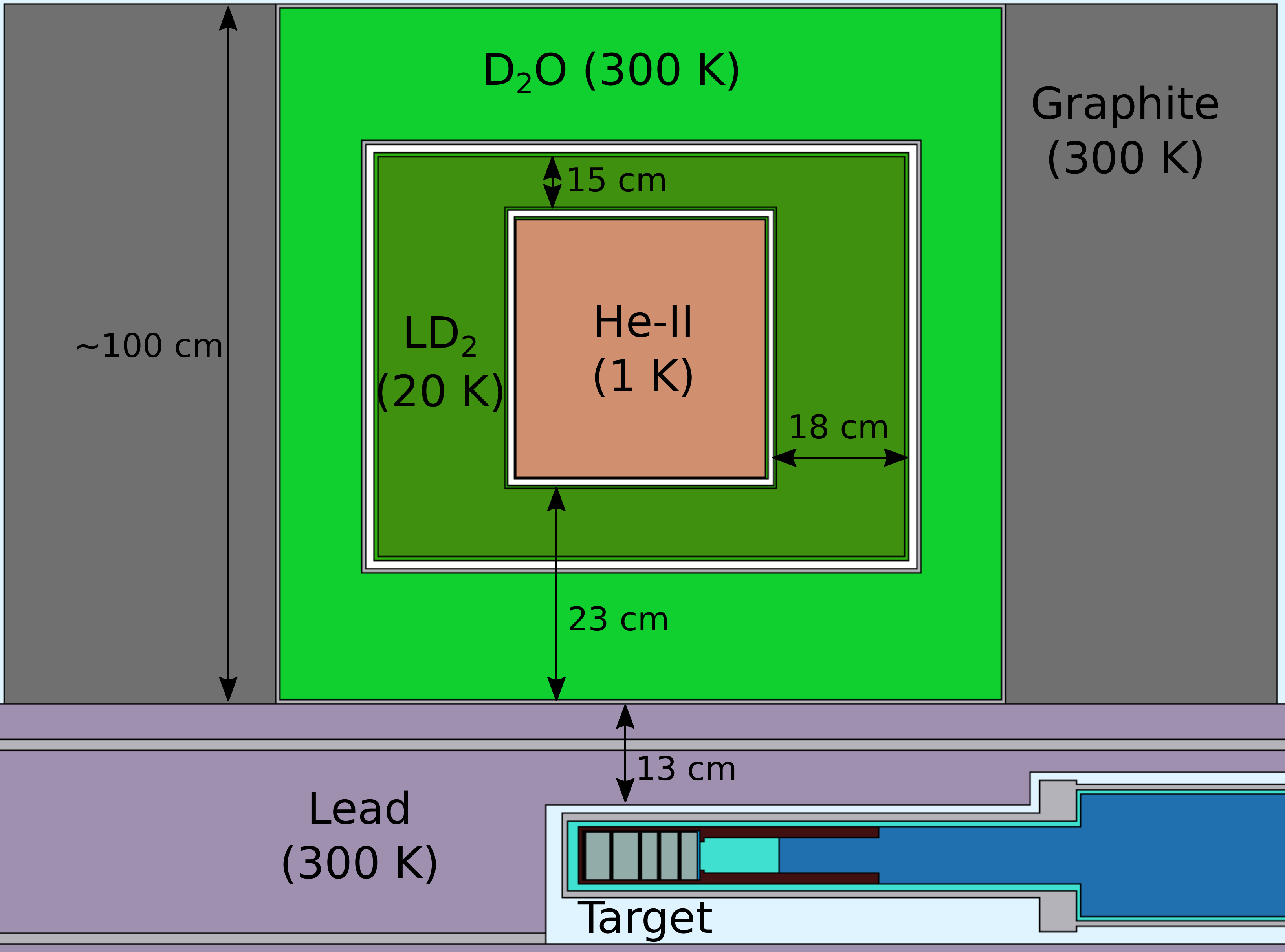}
\includegraphics[height=3.5cm]{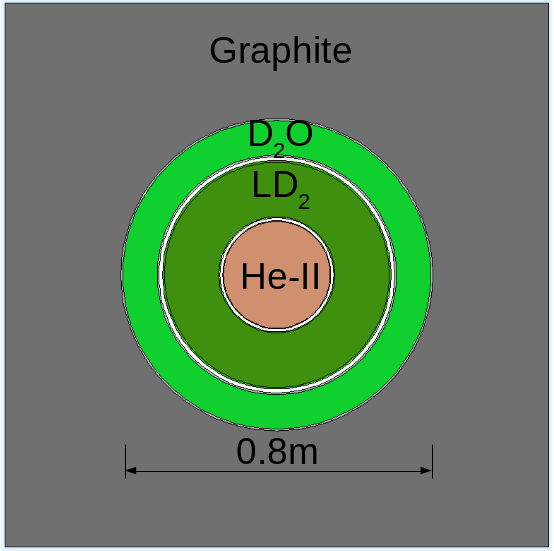}
\caption{Cross sections of the initial moderator geometry viewed from the side (left) and top (right) with the ideal layer thicknesses indicated.}
\label{fig:initialmodel}
\end{figure}

For an initial optimization we chose well-known neutron moderators in a geometry that we assumed makes best use of the symmetry of the problem: concentric, vertical cylinders, centered above the target and filled with heavy water at room temperature, liquid deuterium at \SI{20}{\kelvin} (LD$_2$), and superfluid helium (He-II), see Fig.~\ref{fig:initialmodel}. Graphite surrounds the whole assembly at the sides and acts as a neutron reflector. The target is encased in lead, which is relatively transparent to fast neutrons and shields much of the gamma radiation from the target.

By individually varying the thickness of bottom, radial, and top layers of thermal and cold moderator and the thickness of the lead layer above the target we were able to determine the optimal thickness for each.

The results showed that the radial liquid-deuterium layer is the most important one, achieving best performance at a thickness of \SI{18}{\centi\meter}, see Fig.~\ref{fig:initialmodel}. However, safety regulations require us to minimize the amount of flammable deuterium needed. The optimization showed that this can be achieved without sacrificing performance by partly replacing the bottom deuterium layer with more heavy water---as long as the combined thickness is \SI{23}{\centi\meter} or more---and reducing the thickness of the top deuterium layer to \SI{15}{\centi\meter}.

The radial and upper heavy-water and graphite layers have less impact. Removing either barely changes the production-to-heat ratio, but larger thicknesses increase production and heat load proportionally, helping to scale to higher heat loads if necessary.

The lead layer between target and moderators should ideally be \SI{13}{\centi\meter} thick, however this can be reduced to again achieve higher production and heat loads while only slightly reducing their ratio.

\subsection{Specific parameter studies}

\subsubsection{Converter-vessel materials}

\begin{table}
\centering
\caption{Effect of different materials (cf. table \ref{tab:materials}) of the superfluid-helium vessel on the production-to-heat ratio, relative to pure aluminium. The wall thickness is roughly scaled with the yield strength of each material.}
\begin{tabular}{l r r}
\toprule
Material & Thickness (mm) & Effect on $P/Q$ (\%) \\
\midrule
Aluminium & 2 & (baseline)\\
Al6061 & 2 & -5 \\
AlBeCast 910 & 3 & +5 \\
AlBeMet 162 & 2 & +50 \\
AZ80 & 2.5 & +40 \\
BerAlCast 310 & 1.5 & -5 \\
Beryllium & 1.5 & +90 \\
Magnox AL80 & 4 & +15 \\
\bottomrule
\end{tabular}
\label{tab:bottle_materials}
\end{table}

Since roughly \SI{50}{\percent} of the heat load on the UCN converter is deposited in the vessel walls (see Fig.~\ref{fig:heatload}) we investigated a range of materials to try to reduce that, see table~\ref{tab:bottle_materials}. The ideal material to use is beryllium, thanks to its high strength, low density, low gamma absorption, and low neutron absorption. However, it is extremely expensive. Other promising materials are magnesium-aluminium alloys like AZ80 and beryllium-aluminium alloys like AlBeMet. Constituents to avoid are e.g.~manganese and silver. These have high neutron absorption and can bring down the performance, e.g.~in the case of BerAlCast.
However, compared to these more exotic materials, aluminium 6061 has much better known strength and radiation resistance, so we chose it as the default material. We are currently considering AlBeMet as a potential future upgrade.

\subsubsection{Effect of deuterium temperature and contaminations}

The deuterium-liquefaction system will operate at a pressure of \SI{117}{\kilo\pascal}, giving melting and boiling points of \SIlist{19;24}{\kelvin}~\cite{CDR2018}. In this temperature range the deuterium density varies from \SIrange{0.18}{0.16}{\gram\per\cubic\centi\meter} with the former giving a \SI{5}{\percent} higher production-to-heat ratio.

Deuterium comes in two spin states, para- and ortho-deuterium. Ortho-deuterium offers better neutron moderation. In thermal equilibrium the fraction of para-deuterium is between \SI{2}{\percent} at \SI{20}{\kelvin} and \SI{33}{\percent} at room temperature \cite{mishimathesis}. To determine its effect on  the performance of the source, we replaced the ortho-deuterium used for the optimization so far with \SI{100}{\percent} para-deuterium. This caused a drop in production-to-heat ratio by \SI{20}{\percent}. Since the actual fraction of para-deuterium is at least three times lower, we expect an at least three times smaller reduction of the production-to-heat ratio.

Para-ortho converters are available to reduce the fraction of para-deuterium in the supplied deuterium gas while it is liquefied. Due to the expected small effect on performance, we currently consider adding a para-ortho converter to the gas handling system a low priority and a future upgrade.

Hydrogen contamination in the deuterium gas, however, is a bigger issue. Each percent of hydrogen contamination would cause a \SI{4}{\percent} reduction in the production-to-heat ratio. Isotopic purity of commercially available deuterium is typically quoted as \SI{99.8}{\percent}~\cite{praxairD2}.

\subsubsection{UCN extraction}

\begin{figure}
\centering
\includegraphics[width=0.75\columnwidth]{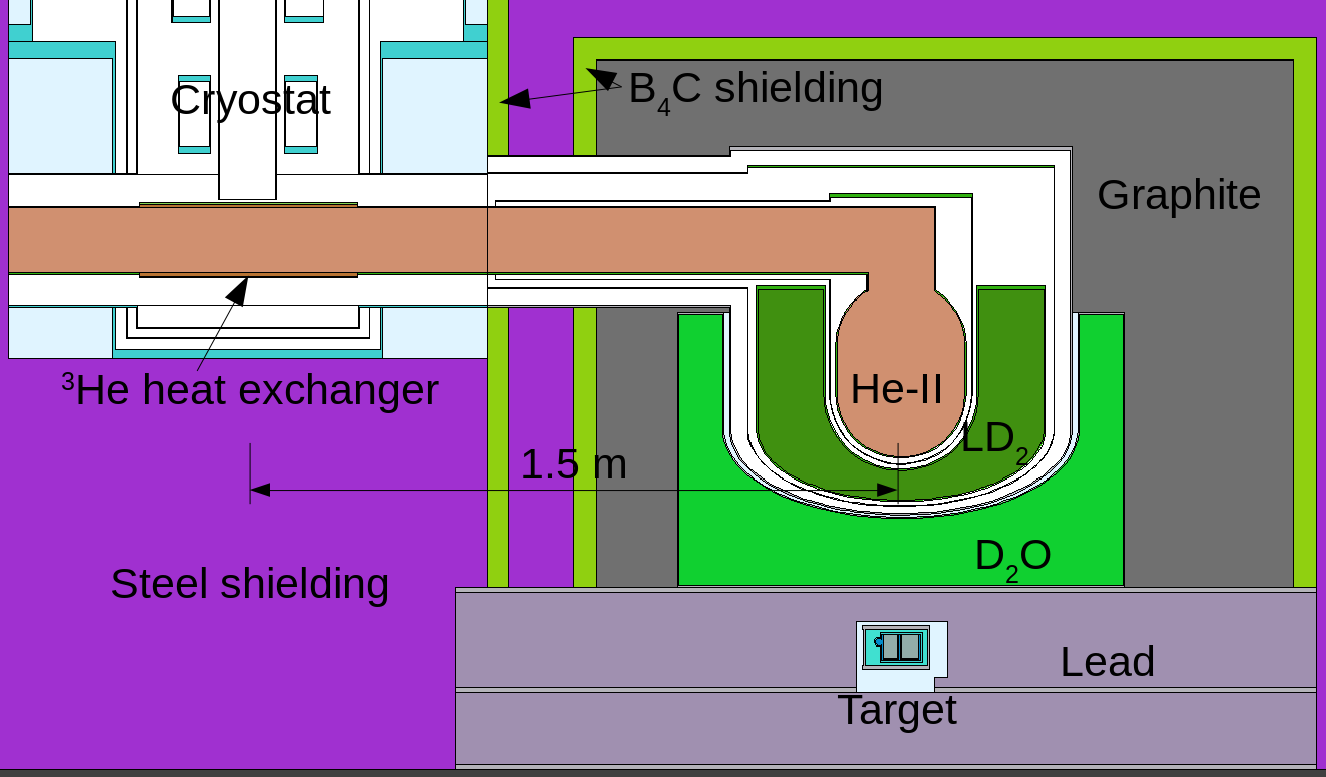}
\includegraphics[width=0.75\columnwidth]{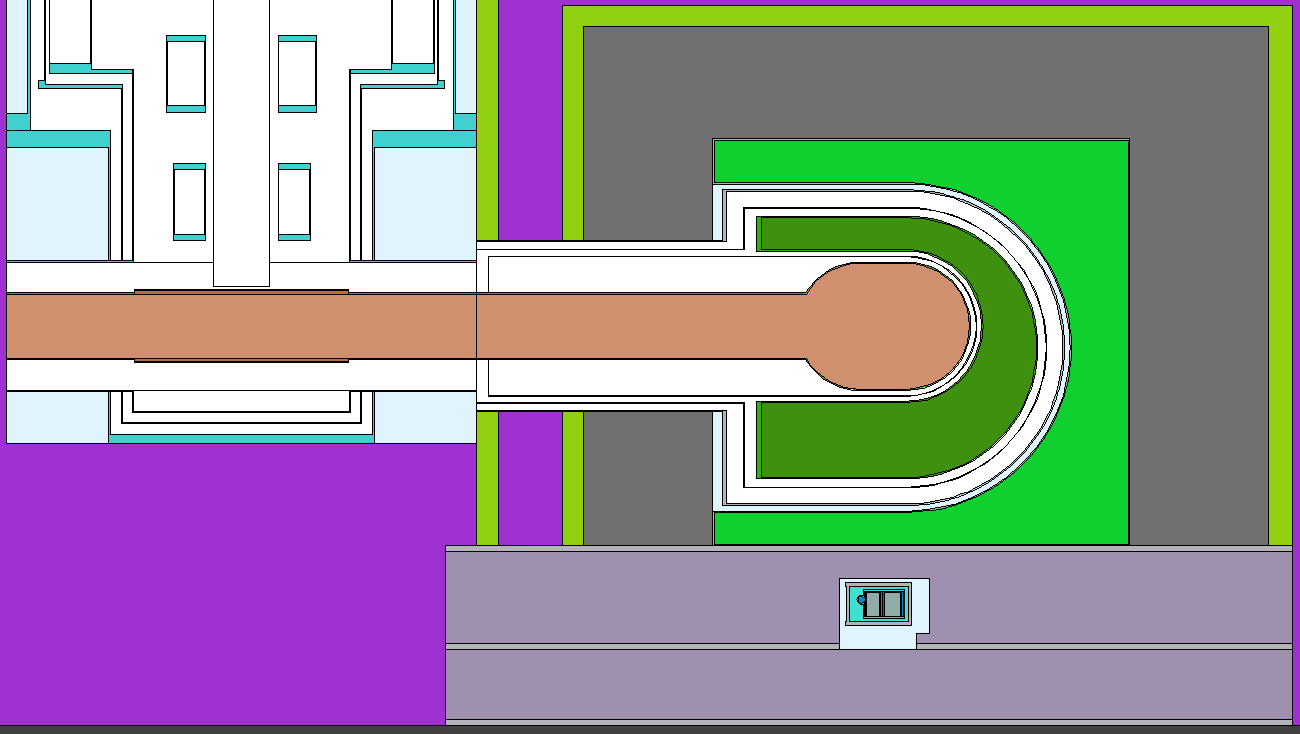}
\caption{Cross section of more realistic simulation models with two different options to extract UCN from the converter through a UCN guide. \textit{Top}: vertical extraction. \textit{Bottom}: horizontal extraction. The distance between converter and heat exchanger was later increased to \SI{2.5}{\meter} to better shield the cryostat from radiation.}
\label{fig:extractionoptions}
\end{figure}

Based on the ideal moderator thicknesses determined in the initial parameter studies, we implemented more detailed geometries, including realistic shapes of the pressure vessels, realistic vessel thicknesses, an additional explosion-proof wall between the deuterium and UCN-converter vessel, UCN guides connecting the converter to a cryostat, a $^4$He-$^3$He heat exchanger, and shielding materials, see Fig.~\ref{fig:extractionoptions}.

The production-to-heat ratio differed by less than \SI{1}{\percent} between the two options, so we chose the horizontal extraction as it is the mechanically more feasible.

\subsubsection{Other cold moderators}

\begin{table}
\centering
\caption{Effect of different cold moderators on the production-to-heat ratio in individually optimized geometries.}
\begin{tabular}{lrrr}
\toprule
& Average layer & & Effect on \\
Moderator & thickness (cm) & Volume (L) & $P/Q$ (\%) \\
\midrule
Ortho-LD$_2$ & 12.5 & 125 & +160 \\
Ortho-LD$_2$ & 19.4 & 200 & +230 \\
Solid D$_2$O & 11.6 & 95 & (baseline) \\
Para-LH$_2$ & 3.6 & 33 & -15 \\
\bottomrule
\end{tabular}
\label{tab:coldmoderators}
\end{table}

The large quantities of liquid deuterium required for the cold moderator pose a substantial safety risk. Therefore we also considered other cold moderators: solid heavy water, liquid hydrogen, solid methane, and mesitylene.

Solid heavy water has the advantage of being inert and easy to handle. However, measurements of its neutron-moderation properties showed that it cannot moderate neutrons below an effective neutron temperature of about \SI{80}{\kelvin}~\cite{sD2Oeffectivetemperature}, reducing the neutron flux at energies where they can be converted to UCN. The exact scattering properties of solid heavy water are also not very well known---MCNP does not contain a dedicated scattering kernel as it does for liquid (heavy) water, graphite, deuterium, and hydrogen~\cite{CHADWICK20062931}. Instead, we had to use a free-gas model at an effective temperature of \SI{80}{\kelvin}, adding doubts about the validity of the results.

Liquid hydrogen has a larger neutron-scattering cross section, reducing the required layer thickness. Although as flammable as deuterium, it would require less volume and reduce safety challenges. However, hydrogen has a 600 times higher neutron-absorption cross section, drastically reducing the time neutrons spend in the cold moderator and converter, reducing UCN production. The increased absorption rate also increases the heat load on the UCN converter due to the gamma radiation generated during the absorption process.

Solid methane and mesitylene could provide significantly higher cold-neutron flux, but they also rely on their large abundances of hydrogen to moderate neutrons, leading to the same issues as with liquid hydrogen. Deuterated variants could avoid that, however no good neutron-moderation data is available for those. Solid moderators are also prone to sudden releases of accumulated Wigner energy, which can lead to a catastrophic failure of their vessel. Due to these reasons we did not study them further.

Using a multi-parameter optimization, similar to the one described in section~\ref{sec:fullopt} but with a constant converter volume, we were able to find the ideal layer thicknesses for each moderator and do a fair comparison of their performance in the UCN source, see table~\ref{tab:coldmoderators}. Since safety regulations limit the volume of liquid deuterium to \SI{150}{\liter}, we ran the optimization for different volumes of liquid deuterium to estimate how a safety-related limit impacts the performance.

The results show that liquid deuterium can offer \numrange{2.6}{3.3} times higher performance than solid heavy water, and \numrange{3.0}{3.9} times higher performance than liquid hydrogen.

\subsubsection{Bismuth neutron filter}

\begin{figure}
\centering
\includegraphics[width=0.75\columnwidth]{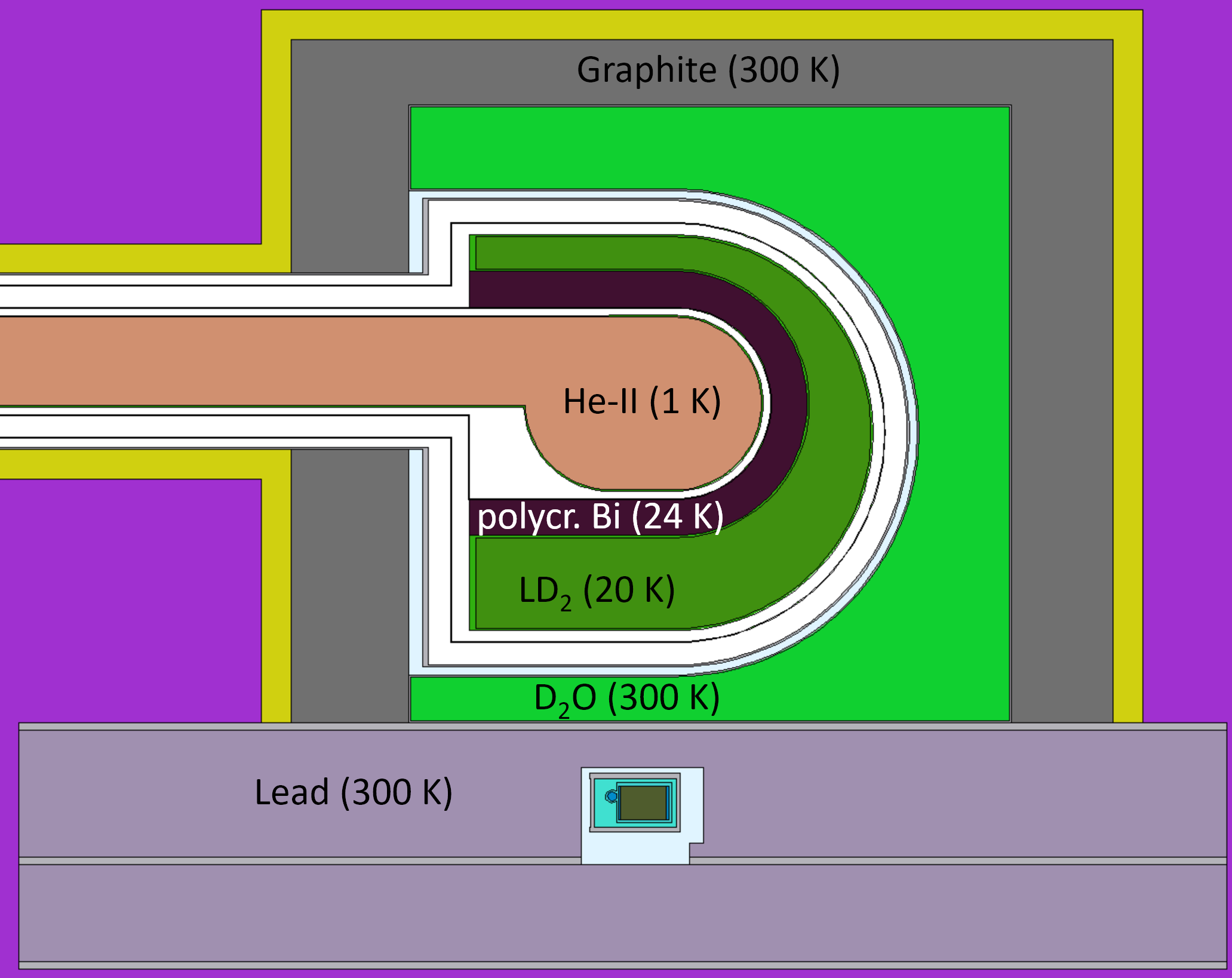}
\caption{Moderator geometry with \SI{5}{\centi\meter}-thick layer of polycrystalline bismuth.}
\label{fig:bismuth}
\end{figure}

Bismuth has a low neutron-absorption cross section and, thanks to its high density, is a good gamma shield. Especially polycrystalline bismuth is often used as a cold-neutron filter since its neutron-scattering cross section sharply drops by at least an order of magnitude for neutrons below an energy of \SI{2}{\milli\electronvolt}~\cite{ADIB200381,bismuth_scatteringkernel}, just above the energy range where the UCN-production cross section peaks.

We tried to make use of these properties with a \SI{5}{\centi\meter}-thick layer of polycrystalline bismuth between the cold moderator and the UCN converter. Ideally, the bismuth would block gamma radiation and thermal neutrons and be almost transparent for cold neutrons, reducing the heat load on the UCN converter.

The simulation showed that such a bismuth layer indeed reduced the heat load to the converter and its vessel by \SI{46}{\percent}. However, the added material and the reduction in LD$_2$-layer thickness to keep its volume below the safety limit also reduced UCN production by \SI{36}{\percent}, resulting in an increase in production-to-heat ratio of \SI{27}{\percent} compared to a configuration without bismuth. Additionally, the bismuth---being coupled to the cold moderator at \SI{20}{\kelvin}---increased the heat load on the cold moderator by almost \SI{200}{\percent} and increased residual radioactivity after long-term operation ten-fold.

\subsection{Full optimization}
\label{sec:fullopt}

Before further optimizing the moderator geometry, we estimated the wall thicknesses required to withstand all overpressure scenarios (e.g.~\SI{3}{\milli\meter} for the converter vessel) and connected the UCN guide to the top of the UCN-converter to minimize the diameter of the penetration through the shielding. And we closed the heavy-water and deuterium vessels around the helium vessel, giving an improvement in estimated UCN density of \SI{15}{\percent}.

\begin{figure}
\centering
\includegraphics[width=\columnwidth]{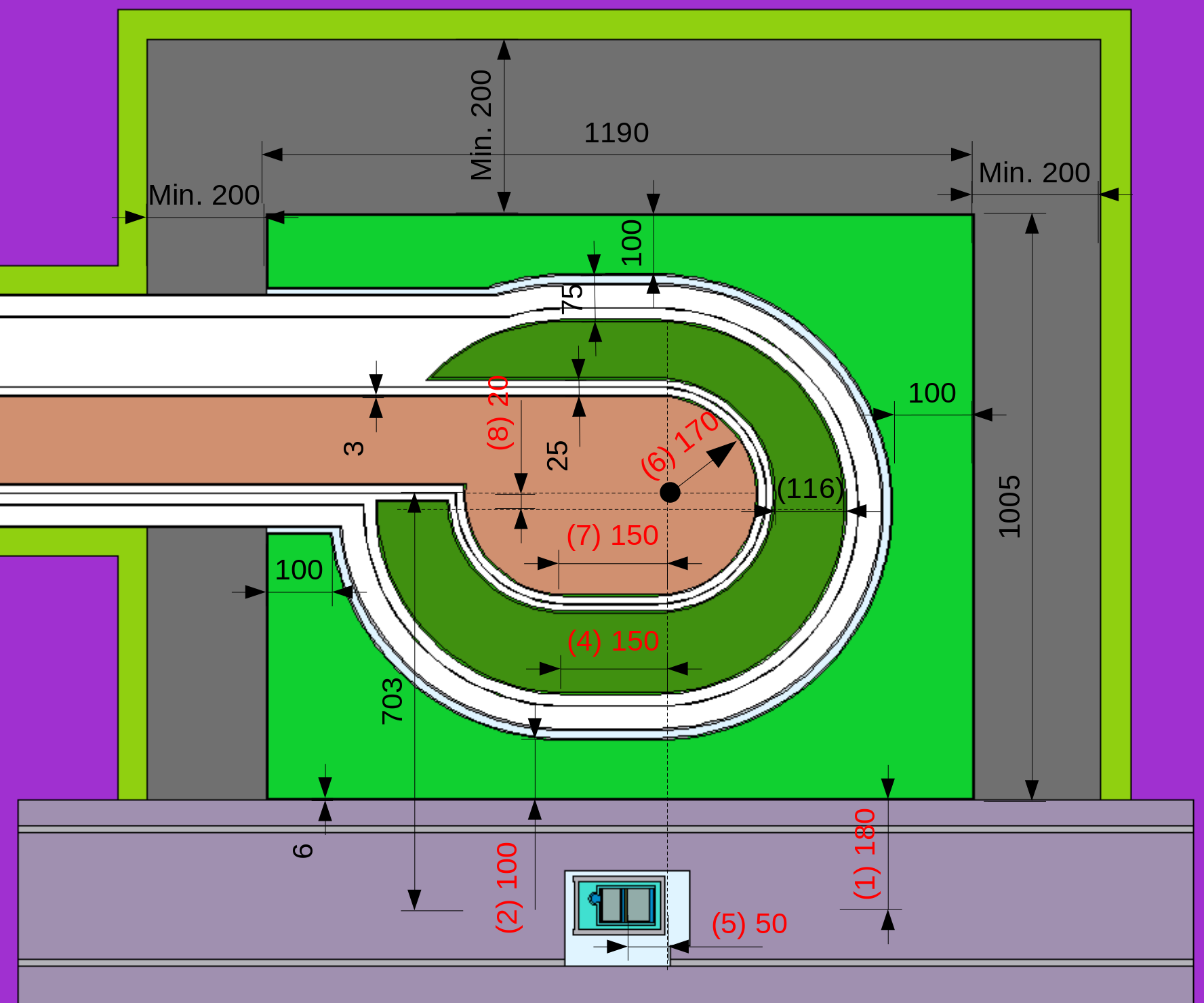}
\caption{Dimensions (all in mm) of an average reference geometry determined from full optimizations with a range of model assumptions. Optimized parameters are indicated in red font. The thickness of the cold moderator (in parentheses) is chosen such that its total volume is 125~L. Vessel walls have a thickness of 3.6~mm, unless otherwise indicated.}
\label{fig:result}
\end{figure}

Then we implemented a multi-parameter optimization, varying eight parameters (see red labels in Fig.~\ref{fig:result}) simultaneously:
\begin{enumerate}[label=(\arabic*)]
\item thickness of lead above target,
\item thickness of heavy-water layer above target,
\item horizontal offset between UCN-converter volume and liquid-deuterium vessel (=\SI{0}{\centi\meter} in Fig.~\ref{fig:result}, not shown),
\item length of the liquid-deuterium vessel,
\item horizontal offset between target and UCN-converter vessel,
\item radius of the UCN-converter vessel,
\item length of the UCN-converter vessel, and
\item vertical offset between liquid-deuterium vessel and UCN-converter vessel,
\end{enumerate}
while adjusting the radius of the liquid-deuterium vessel to keep its volume fixed. After each parameter change the optimization program performs the MCNP simulation and tries to find the set of parameters that maximizes the density estimator $\rho = P \tau / V$ (equation~\ref{eq:densityestimator}).

To keep the total liquid-deuterium inventory below the safety limit of \SI{150}{\liter}, we fixed the cold-moderator volume to \SI{125}{\liter}---taking into account up to \SI{25}{\liter} of additional liquid in piping and heat exchangers. We performed the optimization for different combinations of assumptions for the parameters in table~\ref{tab:assumptions}. From this set of optimized geometries we chose an ``average'' geometry (see Fig.~\ref{fig:result}). This average geometry turned out to be very robust---its performance differed by less than \SI{2}{\percent} from the set of optimized geometries.

\subsection{Methodology}

We modeled the simulation geometries in Flair~\cite{vlachoudis2009flair}, a GUI for the particle-transport simulation Fluka~\cite{Ferrari:2005zk}, and then exported them to MCNP~\cite{doi:10.13182/NT11-135}. The optimization parameters were adjusted by a custom Python program and then optimized using the SLSQP algorithm from the SciPy.optimize library~\cite{Kraft1988}. The MCNP simulations were performed on Compute Canada computing clusters. Git and GitHub proved to be valuable tools to manage the large number of simulation models accumulated during the optimization.

\subsection{Most important parameters}

\begin{table}
\centering
\caption{Possible improvements of the moderator geometry ranked by their potential performance increase.}
\begin{tabular}{lr}
\toprule
 & Increase in \\
Improvement & UCN density \\
\midrule
Beryllium converter vessel & \SI{90}{\percent} \\
AlBeMet or AZ80 converter vessel & \SIrange{30}{50}{\percent} \\
Thinner converter vessel & \SI{25}{\percent} per mm \\
Increased LD$_2$ volume & \SI{5}{\percent} per \SI{15}{\liter} \\
Thinner vacuum separator/LD$_2$ vessel & \SI{5}{\percent} per mm \\
Thinner thermal shield/D$_2$O vessel & \SI{2.5}{\percent} per mm \\
Reduced spacing around converter & \SI{0.5}{\percent} per mm \\
\bottomrule
\end{tabular}
\label{tab:improvements}
\end{table}

The performance of the source is rather robust against changes of the vessel sizes. Multiple iterations of the optimization resulted in differences of a few centimeters, but the estimated density varied by less than \SI{2}{\percent}. The exception is the cold moderator. Adding \SI{15}{\liter} of liquid deuterium increases UCN production by 23~\% and heat load by 14~\%. See table \ref{tab:improvements} for the resulting effect on estimated UCN density.

Much more critical are thicknesses of vessel walls. Especially the UCN-converter vessel directly contributes to the heat load, making up about \SI{50}{\percent} of the total heat load on the converter. Reducing its thickness or changing its material has the largest impact on the performance of the source, e.g.~a converter vessel made of AlBeMet can increase UCN production by \SI{18}{\percent} and reduce heat load by \SI{25}{\percent}. AZ80 magnesium alloy can increase UCN production by \SI{9}{\percent} and reduce heat load by \SI{33}{\percent}. The effect on UCN density in the EDM cell depends on the assumptions, see the estimated range of improvement in table \ref{tab:improvements}.

The other vessel walls also absorb neutrons, reducing UCN-production rate. This effect is most important for the walls between the cold moderator and converter, where an increase of \SI{1}{\milli\meter} in thickness of any of those walls causes a drop in UCN production by \SI{5}{\percent}. The walls between the thermal and cold moderators have about half as much impact since they are in a region with higher-energy neutrons.

The spacing between the converter vessel and the cold-moderator vessel also has slight impact on UCN production. Increasing the spacing by \SI{10}{\milli\meter} reduces UCN production by \SI{5}{\percent}.

\section{Result}

\begin{figure}
\centering
\includegraphics[width=\columnwidth]{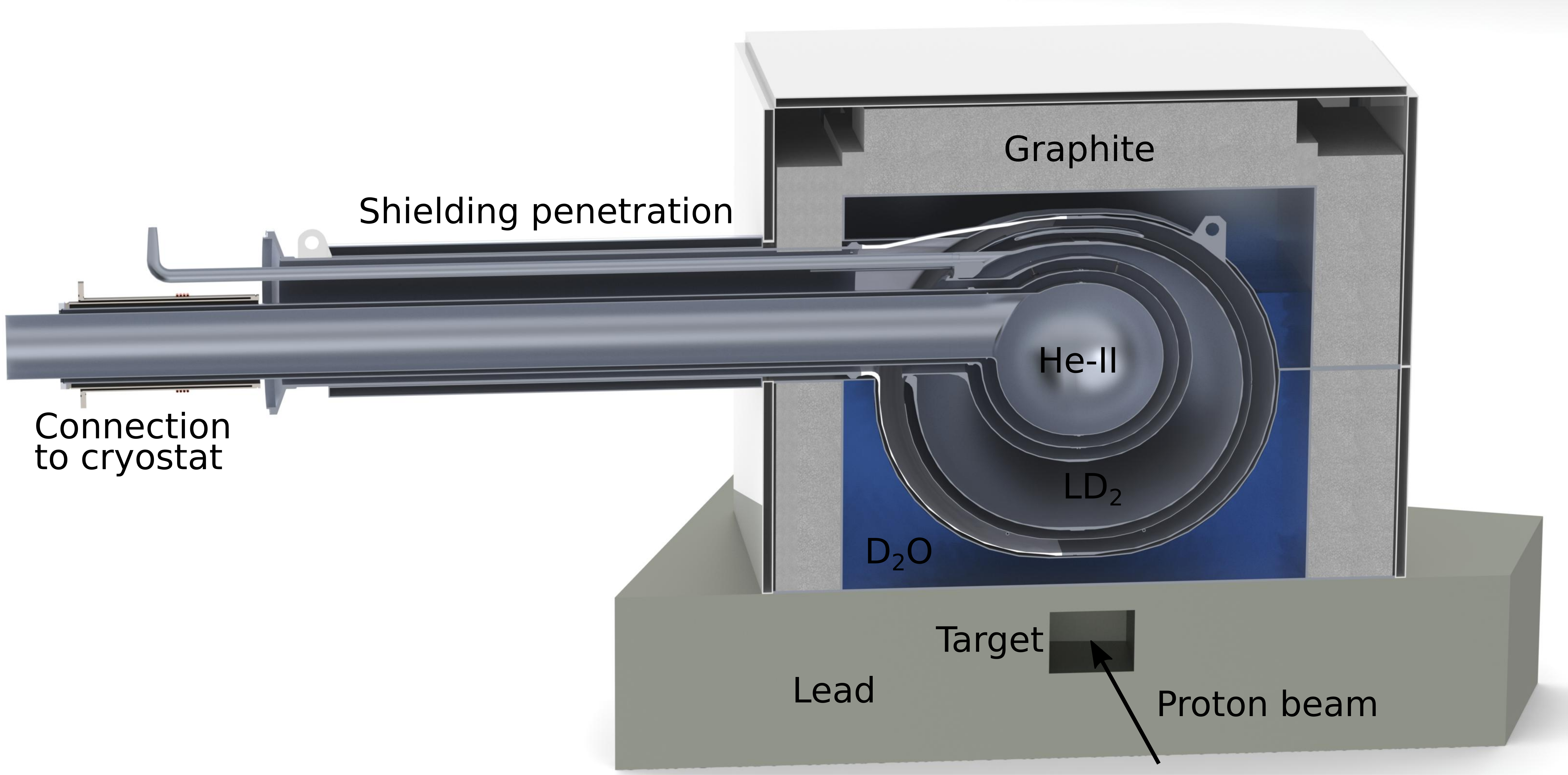}
\includegraphics[width=\columnwidth]{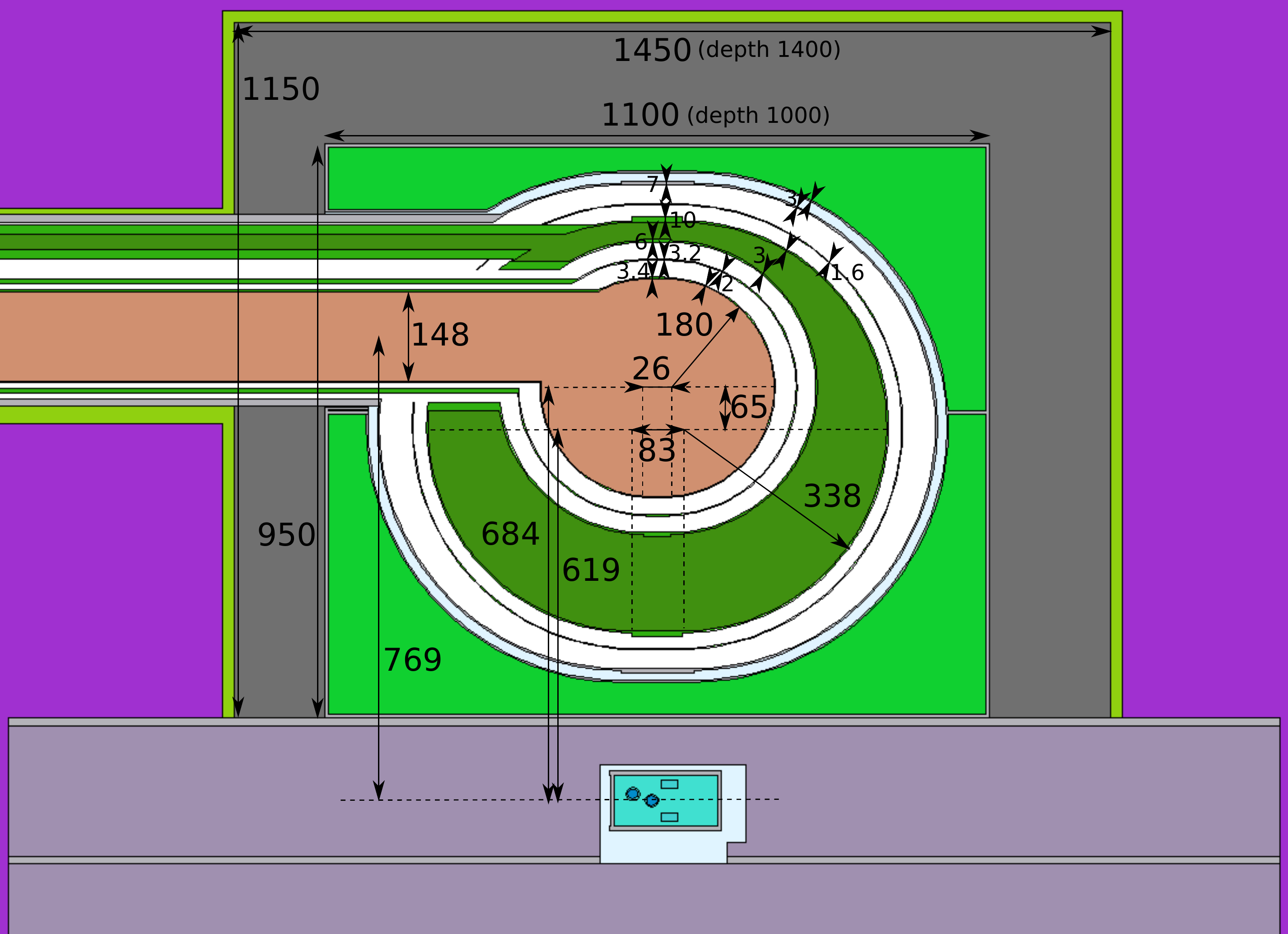}
\caption{Detailed concept for the moderator vessels and reflectors (top) and the corresponding simulation model (bottom). The proton beam points into the page.}
\label{fig:spherical}
\end{figure}

The fully optimized geometry shown in Fig.~\ref{fig:result} served as a basis for detailed engineering of the pressure vessels. Detailed stress simulations with ANSYS to minimize the wall thicknesses showed that the liquid-deuterium vessel would have to have thicker walls to withstand a potential hydrogen explosion, while the converter vessel could be thinned. Especially the welds connecting the hemispherical and straight sections weaken the temper of the aluminium alloy, requiring a locally thicker wall. All changes in vessel sizes, wall thicknesses, and materials were implemented in the simulation to check their impact on performance, and the geometry was re-optimized if necessary. With these engineering restrictions, the optimization favored a stronger aluminium 2219 alloy and almost spherical vessel shapes, since they allow to connect the two hemispherical shells with a single weld, see Fig.~\ref{fig:spherical}.

We also used a similar optimization to minimize the volume and therefore cost of the heavy-water and graphite reflectors, while keeping the drop in performance to less than \SI{5}{\percent}. The minimum amounts required were \SI{630}{\liter} of heavy water and at least \SI{15}{\centi\meter} of graphite on all sides of the heavy-water vessel.

UCN production and heat load are maximized if the vessels are placed right above the target. However, UCN density is maximized if they are moved upstream and left of the proton beam (out of the page and left in Fig.~\ref{fig:spherical}), most likely due to the asymmetry introduced by the UCN guide. The optimal offset can reach \SI{50}{\centi\meter}, depending on the assumptions (cf. table~\ref{tab:assumptions}), but the shielding arrangement allows only small offsets of about \SI{10}{\centi\meter}.

\begin{table}
\centering
\caption{Volumes, maximum heat loads, and heat loads averaged over a duty cycle of \SI{25}{\percent} with the detailed engineering model. The UCN-converter volume does not include the conduction channel.}
\begin{tabular}{lrrr}
\toprule
 & & \multicolumn{2}{c}{Heat load (W)} \\
 & Volume (L) & max. & average \\
\midrule
UCN converter & 27 & 8.1 & 2.8 \\
Liquid deuterium & 125 & 63 & 21 \\
Heavy water & 630 & 430 & 150 \\
\bottomrule
\end{tabular}
\label{tab:results}
\end{table}

\begin{figure}
\centering
\includegraphics[width=\columnwidth]{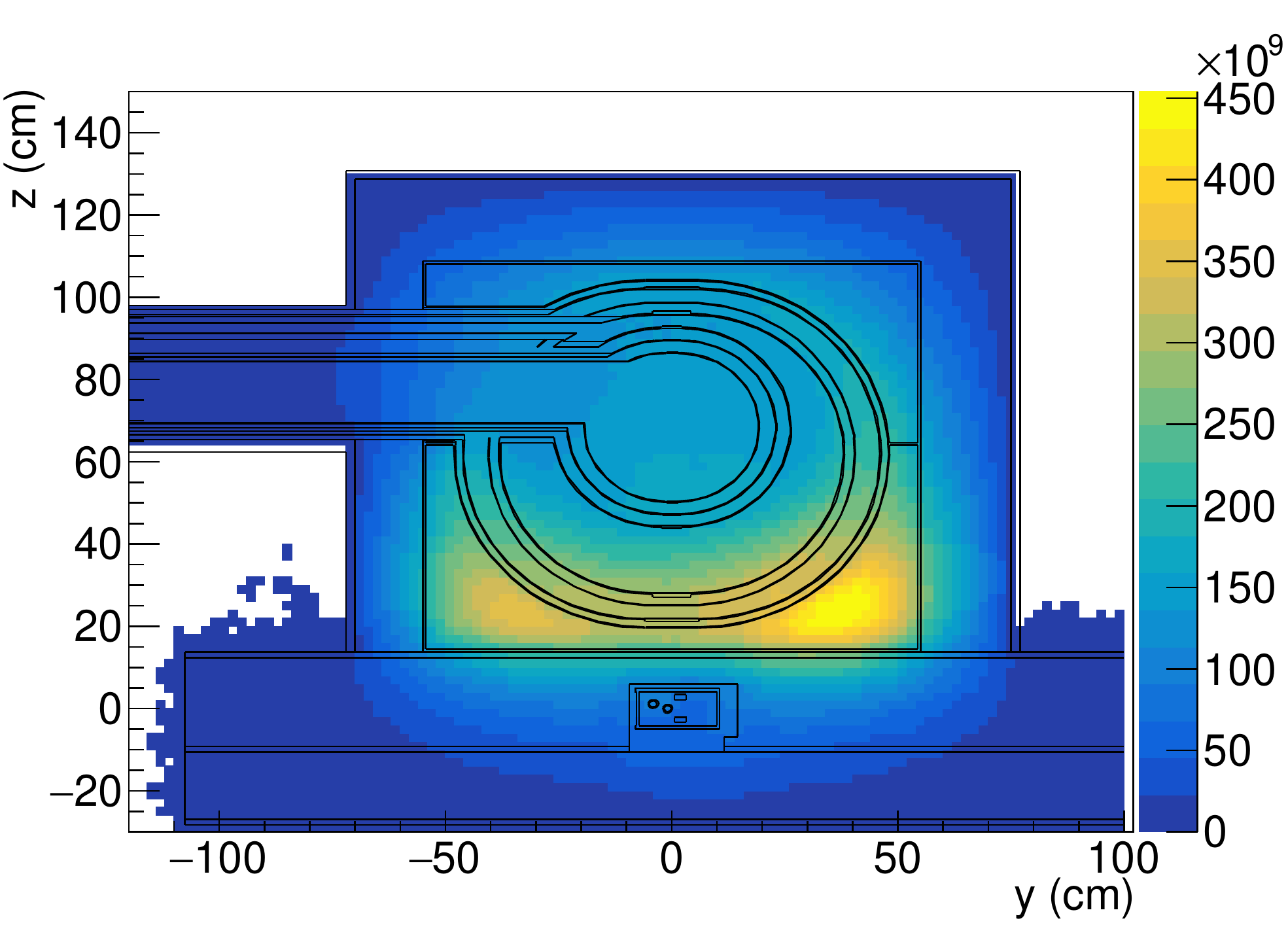}
\includegraphics[width=\columnwidth]{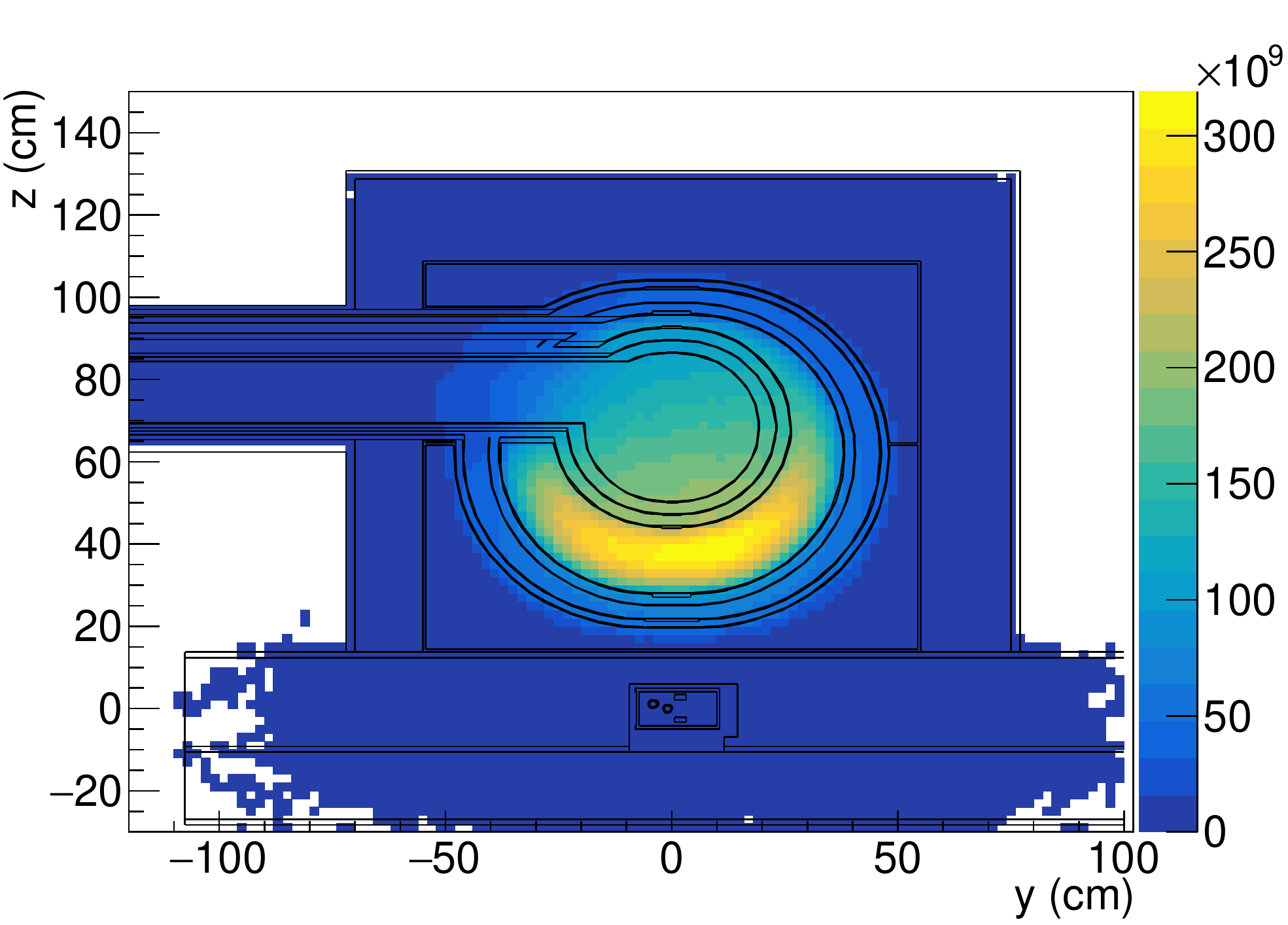}
\includegraphics[width=\columnwidth]{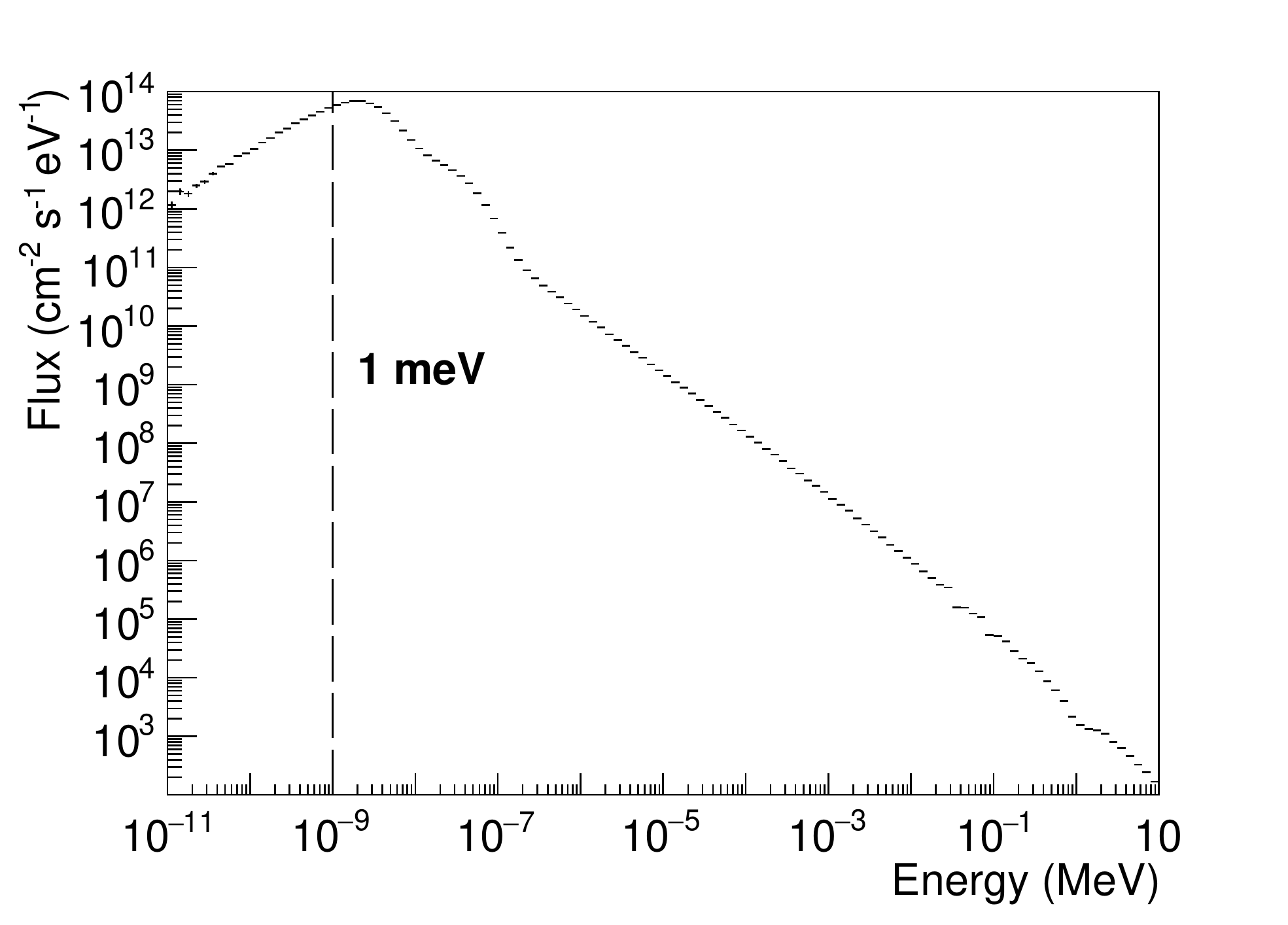}
\caption{Neutron flux per \si{\square\centi\meter} and \si{\second} for thermal neutrons ($\SI{6}{\milli\electronvolt} < E < \SI{100}{\milli\electronvolt}$, top) and cold neutrons ($E < \SI{6}{\milli\electronvolt}$, middle); and the neutron spectrum in the converter (bottom).}
\label{fig:flux}
\end{figure}

With this more detailed geometry we expect UCN production of \SI{1.6e7}{\per\second} and heat loads as shown in table \ref{tab:results}. As expected, the largest fluxes of thermal and cold neutrons are found in the thermal and cold moderators (Fig.~\ref{fig:flux}). The neutron spectrum in the converter has its maximum slightly above \SI{1}{\milli\electronvolt} due to only partial thermalization in the cold moderator.

\cite{leung2019nextgeneration} suggests that using the standard $^4$He-scattering kernel in MCNP overestimates UCN production, since superfluid $^4$He is a less efficient neutron moderator. Following their approach of reducing the $^4$He density by \SI{90}{\percent} to take this into account reduces UCN production by \SI{13}{\percent} to \SI{1.4e7}{\per\second}. Tritium production was also estimated at about \SI{200}{\giga\becquerel} per year when operating at \SI{25}{\percent} duty cycle, with most of it occurring in the heavy-water and liquid-deuterium moderators.

\section{Conclusions}

Thanks to extensive simulations and optimization we found a geometry of neutron moderators and superfluid-helium UCN converter that performs close to optimally for a wide range of assumptions on UCN-storage lifetime in the source, cooling power of the cryostat, and attached experiments with volumes of up to \SI{100}{\liter}. The liquid-deuterium moderator was fixed to a volume of \SI{125}{\liter}, to stay below the safety limit of \SI{150}{\liter}. This geometry served as a baseline for detailed engineering of the pressure vessels, taking into account various safety constraints, and the simulations continued to guide all design decisions.

The final design has a converter volume of \SI{27}{\liter}, has a liquid-deuterium volume of \SI{125}{\liter}, and achieves a simulated UCN production of \SIrange{1.4e7}{1.6e7}{\per\second} at a heat load of \SI{8.1}{\watt}. The most critical parameters are the choice of cold moderator and material of the converter vessel. With a liquid-deuterium moderator the UCN density delivered to a typical UCN-storage experiment is about three times higher than with solid heavy water or liquid hydrogen. AlBeMet, a beryllium-aluminium alloy, is a promising material for the converter vessel and could improve the performance by \SIrange{30}{50}{\percent} compared to aluminium 6061.

The new moderator vessels and source are scheduled to be installed above the spallation target beginning of 2021, with commissioning starting the same year. The projected performance will enable the TUCAN collaboration to measure the neutron electric dipole moment with a sensitivity of \SI{1e-27}{\elementarycharge\centi\meter}.

\section{Acknowledgments}

We would like to thank P.~Carlson and M.~Chin for their early contributions to this work and Y.-S.~Cho for providing an MCNP scattering kernel for polycrystalline bismuth.

This work was supported by the Canada Fund for Innovation (CFI), the Natural Sciences and Engineering Research Council of Canada (NSERC), and Compute Canada.

\section{References}

\bibliography{bib}

\end{document}